# The population doctrine in cognitive neuroscience

R. Becket Ebitz[1]* and Benjamin Y. Hayden[2]

[1] Department of Neuroscience
Faculty of Medicine
Université de Montréal
Montréal, QC Canada

[2] Department of Neuroscience,
Center for Magnetic Resonance Research, and
Center for Neuroengineering
University of Minnesota
Minneapolis, MN USA

**\*Corresponding author:**
    R. Becket Ebitz
    Department of Neuroscience
    Faculty of Medicine
    Université de Montréal
    Montréal, QC Canada
    Email address: becket@ebitzlab.com

**SUMMARY**

A major shift is happening within neurophysiology: a population doctrine is drawing level with the single-neuron doctrine that has long dominated the field. Population-level ideas have so far had their greatest impact in motor neuroscience, but they hold great promise for resolving open questions in cognition as well. Here, we codify the population doctrine and survey recent work that leverages this view to specifically probe cognition. Our discussion is organized around five core concepts that provide a foundation for population-level thinking: (1) state spaces, (2) manifolds, (3) coding dimensions, (4) subspaces, and (5) dynamics. The work we review illustrates the progress and promise that population-level thinking holds for cognitive neuroscience—for delivering new insight into attention, working memory, decision-making, executive function, learning, and reward processing.

**IN BRIEF**

The population doctrine holds that the fundamental computational unit of the brain is the population. This view holds great promise for resolving open questions in cognition. We discuss five core concepts of population analysis and review relevant papers.

# INTRODUCTION

Cognition gives us the flexibility to selectively attend to a stimulus, hold information in mind, pursue arbitrary goals, implement executive control, or decide between actions by learning about and weighing beliefs about their outcomes. Neurophysiology has revealed much about the neural basis of cognition over the last fifty years, generally by examining responses in one neuron at a time. However, brain areas contain hundreds of millions of neurons (Herculano-Houzel 2009) and somewhere between the scale of single neurons and the gross pooling in EEG or fMRI, neuronal populations produce macroscale phenomena that are poised to link the scale of single neurons to the scale of behavior.

The term *population doctrine* describes the belief that the population, not the neuron, is the fundamental unit of computation (Saxena and Cunningham 2019). This idea is not new: some researchers have always studied how neurons behave collectively, whether that was Hebb's cell assemblies in the 1940's (Hebb 1949) or Georgopolous' population vectors in the 1980s (Georgopoulos, Schwartz, and Kettner 1986). Population ideas have always been influential in theoretical neuroscience, but they have sometimes lain dormant among experimentalists. However, with the development and spread of new technologies for recording from large groups of neurons, we have seen a resurgent interest in population-level thinking. Alongside new hardware, an explosion of new concepts and analyses have come to define the modern, population-level approach to neurophysiology (Yuste 2015; Saxena and Cunningham 2019; Jazayeri and Afraz 2017; Shenoy and Kao 2021; Vyas et al. 2020). Although advances in high-yield neural recordings were critical for the resurgence of these ideas, high-yield recordings are neither necessary nor sufficient nor sufficient to make a population neurophysiology paper. Instead, what defines the field is its object: the neural population. To a population neurophysiologist, neural recordings are not random samples of isolated units, but instead low-dimensional projections of the entire manifold of neural activity (Gallego et al. 2017; Jazayeri and Afraz 2017). Here, we will explain this and other fundamental ideas in population neurophysiology, while illustrating why we believe this approach holds such promise for advancing our understanding of cognition.

## THE STATE SPACE

For a single-unit neurophysiologist, the canonical analysis is a neuron's peristimulus time histogram (PSTH). For a population neurophysiologist, it is a neural population's *state space diagram* (**Figure 1A**). Instead of plotting the firing rate of one neuron against time, the state space diagram plots the activity of each neuron against one or more other neurons. At every moment in time, the population is at some *neural state*: it occupies some point in neuron-dimensional space, or, identically, produces some vector of firing rates across recorded neurons. Time is a function that links neural states together; it turns sequences of neural states (or sets of PSTHs) into *trajectories* through the state space (**Figure 1B**). (Trajectories have interesting implications for thinking about how the brain computes. We will return to them in **Dynamics**.) Especially in cognitive studies, neural states may be called *representations*: a somewhat contentious term meant to highlight the correspondence between a neural state and some percept, memoranda, computation, or behavior (see **Text Box: Representations?**).

Re-casting population activity as a neural state can suggest new hypotheses. As vectors in neuron-dimensional space, neural states both point in some direction, and have some magnitude (**Figure 1A**). Because the direction of a neural state vector is related to the pattern of activity across neurons, it is probably unsurprising that state vector direction encodes object identity in the inferotemporal cortex (IT) (Jaegle, Mehrpour, and Rust 2019; Chang and Tsao 2017). However, it is maybe more surprising that this second feature—neural state magnitude—also matters: it predicts how well objects will be remembered later (Jaegle, Mehrpour, and Rust 2019; Jaegle et al. 2019). When we pause to realize that magnitude is essentially a sum of activity across neurons, this may seem like an obvious analysis, or one that just recapitulates the cell-averaged PSTH. However, this is a sum across *all* neurons—regardless of their tuning properties—and it is taken without normalization. For a single-neuron neurophysiologist, it is essential to normalize the dynamic range of neurons before averaging—this is the only way to produce a clear picture of the response profile of an "average" neuron. However, for a population neurophysiologist, more concerned with the holistic population than the average neuron, differences in the dynamic range between neurons are a critical part of the signal. Because spikes are energetically costly and large-magnitude neural states require more spikes, understanding this aspect of the population code could be particularly interesting in cognitive domains where energetic efficiency is a concern, like in classic, energetically costly conflict signals (Ebitz et al.

2020; Ebitz and Platt 2015) or in understanding why we follow simple rules in lieu of less efficient, but more flexible decision-making processes (Cohen et al. 2021; Ebitz et al. 2020).

Because the state space gives us a spatial view of neural activity, it makes it natural to start to think about spatial relationships between different neural states, to reason about *distances* (**Figure 1A**). There are many ways to measure the distance between neural states, including the Euclidean distance, the angle between state space vectors (which decorrelates distance from any differences in magnitude) and the Mahalanobis distance (which accounts for the covariance structure between neurons). (See Walther et al., 2016, for discussion of distance measures in neural data analysis.) Distance measures have many applications, including for reasoning about how neural states evolve over time. For example, in a slowly changing environment, we might expect neural states to also change slowly over time. However, there can be sudden jumps in neural states across pairs of trials (Karlsson et al. 2012; Durstewitz et al. 2010; Bartolo et al. 2020; Russo et al. 2021; Ebitz et al. 2018; Malagon-Vina et al. 2018). These jumps could reflect cognitive or behavioral discontinuities—sudden changes in our beliefs or policies—and are thus difficult to reconcile with simple cognitive models of sequential decision-making, where information is integrated slowly across multiple trials and behavior changes gradually, rather than suddenly (Bartolo and Averbeck 2020). Instead, jumps may better resonate with hierarchically structured models, where inferences are being made at the level of policies, rather than actions (Collins and Koechlin 2012; Eckstein and Collins 2020).

Measures of distance can also be combined with clustering algorithms from machine learning to examine the similarity between neural responses across task conditions (Kriegeskorte et al. 2008; Hunt et al. 2018), characterize hierarchical relationships between different pieces of information (Kiani et al. 2007; Reber et al. 2019), or examine the variance within or between task conditions—a factor that changes with learning (Thorn et al. 2010) and across goal or belief states (Ebitz et al. 2018). Perhaps the most systematized approach for reasoning about spatial relationships between neural states is Representational Similarity Analysis (RSA), a method first developed in neuroimaging that has been extensively reviewed in that context (Kriegeskorte et al. 2008; Diedrichsen and Kriegeskorte 2017). In neurophysiology, RSA has been used to ask how semantic knowledge shapes mnemonic representations (Reber et al. 2019), to compare the computations involved in decision-making across regions (Hunt et al. 2018), to characterize the stability of representational structure over working memory (Spaak et al. 2017), to examine how rules shape information processing across regions (Ebitz et al. 2020), and to contrast the contributions of different structures to encoding task dimensions (Keene et al. 2016). In fMRI, RSA is often interpreted as a kind of neural recapitulation of perceptual or psychological similarity. However, because we still lack whole-brain resolution, neurophysiologists should be cautious about these kinds of interpretations because different representational structures may be found in different brain regions (Hunt et al. 2018; Keene et al. 2016).

Neural state space diagrams can have as many axes as there are recorded neurons. However, neural activity often only varies along a smaller number of directions in the state space, known as *dimensions* (Gao and Ganguli 2015; Gao et al. 2017; Low et al. 2018; Gallego et al. 2018; 2017; Chaudhuri et al. 2019; Lehky et al. 2014). Because activities of different neurons are correlated with each other (Cohen and Kohn 2011), every neuron does not make an independent contribution to the population (Umakantha et al. 2020). It is generally a combination of neurons that drive most of the variability in neural activity. We take advantage of this redundancy by using *dimensionality reduction* to compress neuron-dimensional state spaces onto a smaller number of axes (Cunningham and Yu 2014). There are many ways to choose these axes, including principal components analysis (PCA; **Figure 1C**). Other dimensionality reduction methods, developed specifically for neurophysiology data, can take advantage of the temporal structure of this data and even solve other problems, like combining together ("stitching") non-simultaneously recorded data sets (Yu et al. 2009; Pandarinath et al. 2018; Kobak et al. 2016; Williams et al. 2018). Because it is hard for humans to reason about higher dimensional spaces, dimensionality reduction is helpful for exploratory analyses, but it can also be an important part of data processing pipelines when we want to focus on only the important dimensions in neural activity.

Is there always a one-to-one correspondence between a neural state and a particular cognitive state? There is some evidence that multiple patterns of activity could implement the same function at different moments in time (Malagon-Vina et al. 2018; Rule et al. 2020; Driscoll et al. 2017; Ebitz et al. 2018). This idea is known as *multiple realizability* in philosophy of mind. For example, in uncertain environments, decision-makers often pass through some period of exploration in between longer periods of following some rule or policy (Wilson et al. 2021; Ebitz et al. 2018 and 2019). Exploration produces the kinds of sudden jumps in neural activity we introduced with the concept of distances, but it also disrupts long-term autocorrelations between neural states and promotes new learning. The pattern of activity that implements a policy after exploration is not the same as the pattern that existed before, even when subjects are just returning to an old policy. The ability to implement the same policy via slightly different neural states could offer some benefits in nonstationary environments (Ajemian et al. 2013; Chambers and Rumpel 2017), but it also implies that the neural state spaces may have many *sloppy* dimensions—dimensions along which neural activity can vary without affecting cognition and behavior—and a smaller number of *stiff* dimensions in which comparatively small differences between neural states can have big implications for cognition and behavior. We will return to these concepts when we discuss **Coding Dimensions** and **Subspaces**.

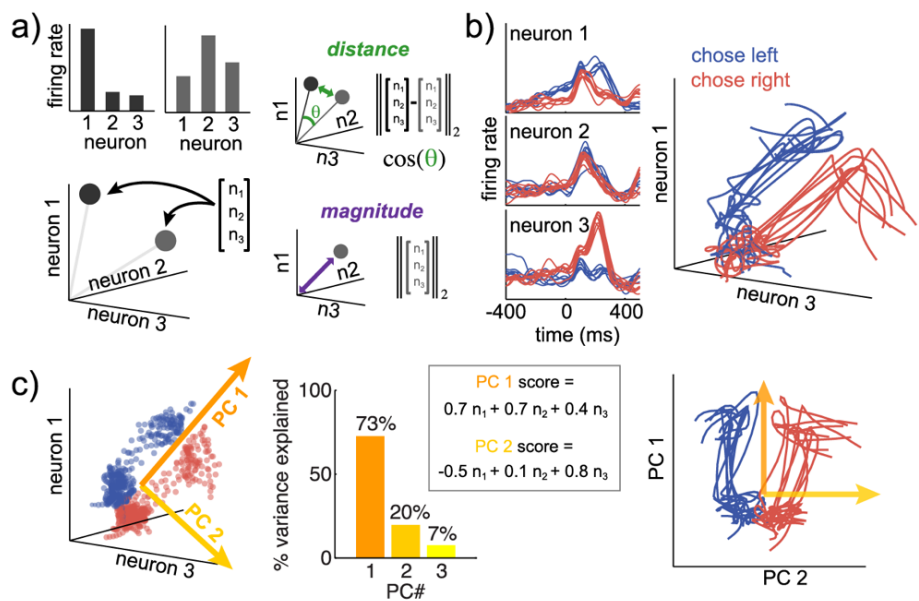

**Figure 1: Neural state spaces and dimensionality reduction**

**A)** A neural state is a pattern of activity across a population of neurons. Neural states can be represented as histograms of firing rates across neurons (top left) or as points or vectors in neuron-dimensional state space (bottom left). The state space representation makes it more natural to think about neural activity geometrically. We can use this as a starting point for reasoning about the distance between different neural states (top right), whether distance is measured via the Euclidean distance, cosine angle, or some other measure. It also illustrates the geometric interpretation of state magnitude (bottom right), which is the neural state's distance from the origin, the length of the neural state vector.

**B)** Peri-stimulus time histograms (PSTHs) plot the average firing rate of single neurons as a function of time, aligned to some event. Spikes are discrete and noisy, so we smooth spike trains by averaging across trials or by smoothing spikes in time. Both were done here to data from (Ebitz et al. 2018) (10-trial averages, gaussian smoothing, σ = 25 ms). We can plot these traces in neural state space, in which case they are called *neural trajectories*: paths linking neural states over time.

**C)** To reduce noise or compress neuron-dimensional state spaces for intuition or visualization, we use dimensionality reduction methods like principal components analysis (PCA). PCA finds an ordered set of *orthogonal* (independent) directions in neural space that explain decreasing amounts of variability in the set of neural states. The first principal component (PC 1) is the direction vector (linear combination of neuronal firing rates) that explains the most variance in neural states (here, 73%). It is often related to time. PC 2 is the direction vector that explains the next most variability, subject to the constraint that it is orthogonal to PC 1, and so on. Right) Projecting neural activity onto a subset of the PCs (here, the first 2) flattens our original 3-dimensional example into a 2-dimensional view that still explains 93% of the variability in neural states.

---

**Text Box 1: REPRESENTATIONS?**

Representation is the process by which one instantiation of some phenomena is replicated in another form (Kosuth, 1965; Churchland and Sejnowski 1990; Brette 2019). An apple can be represented in a painting or in a pattern of activity across a group of neurons. It seems almost trivial to use the word "representation" to refer to the neural correlate of a stimulus, memoranda, cognitive process, or action, yet the word is contentious. Why?

For motor neurophysiologists, the term "representation" is loaded, in part, by the historical influence of sensory coding frameworks in this domain. Rather than inverting the sensory-coding model to focus on how neural activity produces movement, it can be tempting to look for "representations" of movement parameters (Vyas et al. 2020; Fetz 1992; Michaels, Dann, and Scherberger 2016). However, while a representation of sensory information may give us some insight into the processes that shaped it—some features may be emphasized, others eliminated, for example—finding a neural representation of movement velocity cannot tell us how high velocity movements will be generated (Cisek 2006).

We have introduced representations as *pictorial* phenomena—as an image of some neural process—but the term representation can also have *mechanistic* connotations. We tend to think of representations as acting, much like a symbol, to hold information, manipulate it, or pass it between regions (Barack and Krakauer 2021), but this use of the term should always be met with two questions: 1) who is interpreting these symbols and 2) how are they doing so?

> This mechanistic representationalism is likely inherited from cognitive science, a field where mental representations are mechanisms for thought: structures on which computations act. However, despite the similarity in language, a neural representation is not the same thing as a mental representation and the former is not evidence for the latter (Baker et al. 2021; Barack and Krakauer 2021). Consider value-based decision-making. Many regions represent value, in the sense that we can decode value information from them (Vickery, Chun, and Lee 2011; Hunt and Hayden, 2017). We could conclude that these regions function to represent value for other regions to act on. However, value representations do not need to have representational functions. It is just as probable that they are a byproduct of the computations needed to execute choices, which happen to be related to the quantity we calculate as value (Hayden and Niv 2021; Fine and Hayden, 2021).
>
> Avoiding the term "representation" leaves us using convoluted language to describe patterns of neural activity that correspond to some event. It is not incorrect, in our view, to say that part of a neural state space represents a task demand or that a neural trajectory represents a sensorimotor transformation, provided we remember that representation is not necessarily the function of that pattern of activity.

## THE MANIFOLD

Because activity of neurons tends to be correlated with each other, because the wiring between neurons constrains what patterns of neural activity are possible, neural states often only vary along a small number of dimensions in the neural subspace. To put it another way, there is a lot of white space in our state space diagrams: neural activity tends to occupy fewer neural states than it would if each neuron made an independent, random contribution to population activity. The part of the neural state space that contains the states that we observe is called the *neural manifold* (**Figure 2A**).

We have at least two notions of a manifold. The first—the one we referred to when we said that neural recordings are a low-dimensional projection of an entire manifold of neural activity (Gallego et al. 2017)—might be better called the "Manifold": this is the space that encompasses all the states that are possible, the states we would observe if we could record forever, from all the neurons. However, another common usage refers to the space containing on-task neural states recorded from a small number of neurons during a finite number of trials. This subtle distinction is why there is no guarantee that a manifold will be the same across tasks, states, or long periods of time. In practice, they often are (Gallego et al. 2018; 2020; Chaudhuri et al. 2019), though it may be possible to learn to generate "new" neural states, given sufficient time and experience (Oby et al. 2019). The difference between the manifold and the Manifold is also why it is meaningful to say that it is easier to learn to generate neural states that are on the manifold than states that are in the white space (Sadtler et al. 2014; Oby et al. 2019): off-manifold neural states may not be impossible, so much as rare. (Of course, "manifold" is also just a generic mathematical term, so it occasionally appears in other contexts, like to refer to the space containing the neural states that encode some piece of information (Sohn et al. 2019; Okazawa et al. 2021). We will return to this in **Coding Dimensions**.)

Because manifolds are spaces, they have geometric properties, including a *dimensionality*—meaning the number of dimensions that are needed to describe them. Different dimensionalities might be better for different computations (**Figure 2B**). Theoretical work suggests that the dimensionality of neural activity could be yoked to the dimensionality of the task (Gao et al. 2017; Gao and Ganguli 2015). Though it is a bit harder to think about tasks geometrically, they can, like manifolds, have an intrinsic dimensionality. If we choose to move left or right based on a single piece of information (like the direction of moving dots), we only need a 1-dimensional manifold to represent the evidence for left-or-right. If it matters that we decide at a particular time, we may need a second dimension to keep track of time, but the critical point is that laboratory tasks are often low-dimensional.

Although the number of dimensions is generally obvious in motor tasks, this is not always true of cognitive tasks (Akrami et al. 2018; Chen et al. 2020; Constantinople et al. 2019; Filipowicz et al. 2020; Glaze et al. 2018). For example, mice, when given an essentially 1-dimensional cognitive task (choosing between 2 images), made decisions that depended on completely irrelevant information (like image locations, past rewards, and past choices (Chen et al. 2020)). Behavior had many more dimensions than the task. This is neither specific to that task (Akrami et al. 2018; Constantinople et al. 2019) nor some artifact of the difficulty of designing good tasks for rodents. Humans also consistently overestimate the dimensionality of cognitive tasks (Glaze et al. 2018; Filipowicz et al. 2020). We would argue that the brain is generally also engaged in task-irrelevant processing, and any extra-task information would also increase manifold dimensionality beyond the dimensionality of the task. Throughout the brain, we find signals that fluctuate with arousal (Ebitz and Platt 2015; Vinck et al. 2015; Engel et al. 2016; Engel and Steinmetz 2019) or irrelevant movements (Musall et al. 2019; Vinck et al. 2015). Indeed, there is some empirical evidence that manifolds can be higher dimensional than the task (Low et al. 2018), though more work is needed to understand how the dimensionality of neural manifolds is related to the dimensionality of tasks and/or behavior.

Some theoretical work suggests that there may be computational benefits to higher dimensional manifolds (Bernardi et al. 2020). With a higher dimensional manifold, we have more flexibility in how we can decode information from neural activity. As dimensionality increases, new decoding strategies become possible that group together increasingly arbitrary collections of neural states, meaning that a downstream structure could, in theory, use a decoding strategy that groups together apples and penguins, but excludes pears. If executive, prefrontal regions function, in part, to create this flexibility, then they should have higher dimensionality than other regions in the same task. However, few studies have compared neural manifolds across regions (though see Thura et al. 2020; Russo et al. 2020) and others might predict an opposing pattern: a progression from a high dimensional encoding of basic visual features to a low dimensional encoding of task dimensions as you move up the cortical hierarchy (Yoo and Hayden 2018; Lehky et al. 2014; DiCarlo et al. 2012; Pagan et al. 2013). Indeed, one important recent study dealt directly with the question of dimensionality in visual cortex (Stringer et al., 2019). In recordings from around 10,000 neurons, this paper found that coding in visual cortex was quite high dimensional, but that its specific dimensionality depends on the dimensionality of the input. Because visual stimuli are generally much higher dimensional than standard laboratory tasks, this result suggests future studies should take advantage of high dimensional cognitive tasks, such as those using continuous parameter and response spaces (Yoo et al., 2021).

Although manifolds are often consistent across time and task conditions (Gallego et al. 2018; 2020; Chaudhuri et al. 2019), some behaviors may recruit fewer dimensions than others. For example, one influential study found that neural activity had different dimensionality during correct versus error trials (Rigotti et al. 2013). It is hard to measure neural dimensionality, but Rigotti et al. developed a clever approach based on counting the number of linear classifiers that could be trained on the data. More classifiers could successfully classify out-of-sample data on correct trials compared to errors, implying that dimensionality went down during errors.

However, classifiers can fail either because a dimension does not exist or because there is more trial-to-trial variability in neural responses. If the latter explanation is correct, then manifold dimensionality could actually be higher on error trials than correct trials. Indeed, quite a few errors are caused by exploratory processes (Ebitz et al. 2019; Pisupati et al. 2021), which do increase both trial-by-trial variability (Ebitz et al. 2018; Muller et al. 2019) and neural dimensionality (Bartolo et al. 2020; Ebitz et al. 2020).

The neural manifold is a powerful concept for reasoning geometrically about neural activity, but we are only at the precipice of understanding its implications for cognition. We do not yet know why a manifold will have a particular shape and/or dimensionality, how its geometric features are linked to cognition, or how they change across regions and why. Nor do we know what portion of the Manifold we are actually recovering in a few hours of recording in a specific task. Developing new methods to measure neural manifolds experimentally will be key.

## CODING DIMENSIONS

One foundational idea in neurophysiology is that neurons are tuned for (i.e., respond differentially to) stimuli, motor responses, cognitive variables, or combinations thereof. Neural populations can also be tuned: neural states can co-vary with task information along specific directions in the neural state space. The directions that best correspond to—that are "stiff" with respect to—some stimuli, cognitive variables, or combinations thereof are known as *coding dimensions* (**Figure 2C-F**). Coding dimensions are not just an aggregate property of tuned neurons: highly tuned neurons certainly contribute to coding dimensions, but so too do untuned neurons (Leavitt et al. 2017). Coding dimensions are thus an emergent property of a neural population.

Coding dimensions are closely linked to the problem of decoding from a neural population. We find coding dimensions by fitting a model to predict task information from neural data. Decoding from the model, then, often involves a step that projects neural data onto the classification axis in the model. Critically, this step also has a geometric meaning: we are using the coding dimension as a new axis for representing neural activity, much like we did with the PCs in Figure 1. However, here we are re-representing a neural state according to how A-like or B-like the population is, rather than where it falls along one PC or another. When used in this geometric sense, this process is known as *targeted dimensionality reduction* because we are reducing the dimensionality of the population to only consider the axes along which it encodes some piece(s) of information (Cunningham and Yu 2014; Mante et al. 2013).

Coding dimensions and targeted dimensionality reduction are powerful tools for linking neurometric and psychometric functions (**Figure 2D-F**). For example, targeted dimensionality reduction was used to get, for the first time, a trial-by-trial measure of attentional modulations in visual area V4 (Cohen and Maunsell 2010). Variability in each trials' projection along the attention coding dimension predicted accuracy and response time in detecting a subtle stimulus change on single trials. Of course, targeted dimensionality reduction can also be used to ask if trial-by-trial information about one variable (like the value of an option) predicts an entirely different variable (like the choice an animal will make) (McGinty and Lupkin 2021)—linking neural computations to the behavior thought to depend on those computations. As the number of simultaneously collected neurons has continued to grow, so has the temporal precision we can achieve with these methods (**Figure 2G-I; Text Box**: **Single Trial Resolution**). One recent study was even able to identify momentary fluctuations in decision variables in real time (Peixoto et al. 2021). This study is important not only because it set a new bar for the temporal precision of decoding, but also because of its unique, closed loop approach. By triggering pulses of information according to the instantaneous value of the coding dimension projection, this study was able to show that these signals are not simply some statistical trick, but, instead, meaningful indices of ongoing computations.

Coding dimensions need not be static, either within (Spaak et al. 2017; Stokes et al., 2013 and 2015; Lin et al. 2020; Kimmel et al. 2020; Cavanagh et al. 2018) or across trials (Malagon-Vina et al. 2018; Rule et al. 2020; Driscoll et al. 2017; Ebitz et al. 2018). Across trials, we have already noted that multiple realizations of the same goal or belief state may be implemented by slightly different patterns of activity (Malagon-Vina et al. 2018; Rule et al. 2020; Driscoll et al. 2017; Ebitz et al., 2018). However, coding dimensions can also be dynamic within trials. For example, in working memory, sustained activity in delay neurons was once thought to be responsible for maintaining items in working memory. However, if delay neurons were solely responsible for maintaining memoranda, then population coding would just be some consistent linear combination of delay neurons; it would not change over time. However, coding dimensions often do change over the course of a memory period (Spaak et al. 2017; Cavanagh et al. 2018; Stokes 2015). Dynamic codes could emerge from the need to keep track of time or reflect a transformation of sensory information into motor preparatory signals. However, they are not inevitable: other studies report stable population codes within similar task epochs (Kimmel et al. 2020; Murray et al. 2017).

What can we learn from coding dimension geometry? Some researchers are working to understand geometric features of coding dimensions, like their curvature (Sohn et al. 2019; Okazawa et al. 2021; Thura et al. 2020). In this context, the term manifold reappears, used now to refer to the (often nonlinear) shape of neural states that covary with some task variable. There is growing evidence that some linear variables may actually be encoded along a curved surface in neural state space (Sohn et al. 2019; Okazawa et al. 2021). It remains unclear if curvature facilitates a particular readout (Sohn et al. 2019), encodes some variable in its own right, or is an epiphenomenal consequence of constraints on firing rates (Okazawa et al. 2021). In any case, these results presage a future in which understanding the shape of coding dimensions could provide insight into cognitive functions.

Future studies may be able to use the relative geometry of different coding dimensions to understand cognition. For example, looking at a recurrent neural network (RNN) trained to perform a variety of tasks, a recent study found striking alignment between coding dimensions that corresponded to specific cognitive functions across tasks (Yang et al. 2019). Population activity was displaced in the same direction in the neural state space whenever the task required an item to be held in working memory, for example. This implies that working memory was executed by a particular coding dimension that was reused across tasks. Of course, experimental work is needed to determine if the brain can compose either coding dimensions or cognitive functions, but this study illustrates two important points. First, that coding dimensions could be the key to answering big questions in cognitive neuroscience, like whether cognitive functions are composable or even unitary. Second, that theoretical work is critical for generating new hypotheses and neural population analyses.

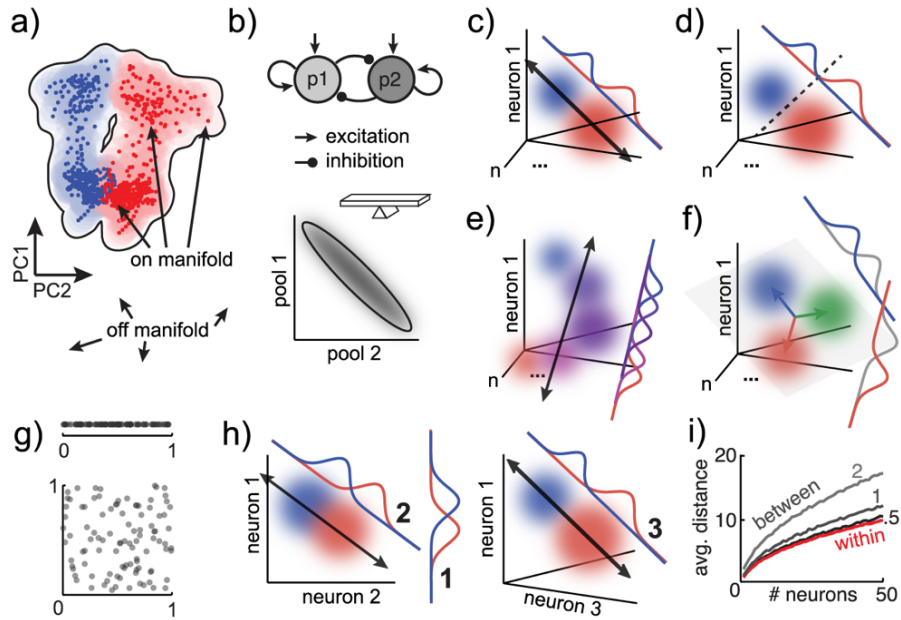

**Figure 2: Manifolds and coding dimensions**

**A)** A toy manifold for the data from Figure 1, illustrating some on- and off-manifold states.

**B)** In a system with two pools of mutually inhibitory, self-excitatory neurons, the manifold would be an almost 1-dimensional negative correlation between the two neurons. This is sufficient for any computation that a balance beam could perform, like measuring the difference between 2 inputs.

**C-D)** Coding dimensions are direction vectors in a state space that explain variability across task conditions (here illustrated as colored Gaussian distributions). With 2 task conditions, coding dimensions can be identified via linear (C) or logistic regression (D). Linear regression fits a line that connects the two states (black arrow), so we decode by projecting data onto the regression line (red and blue distributions). Logistic regression finds a classifier that discriminates the two states, so the distance from the separating boundary is the decoding axis.

**E)** When there is a continuum of conditions, we can use linear regression to identify a classifier, even when the states are arranged non-linearly. A linear approximation captures most of the variance in most curved functions and, at least in some circumstances, behavior may itself reflect a linear readout from a curved representation (Sohn et al. 2019).

**F)** When there are more than 2 conditions, multiple-class models can identify a set of coding dimensions: a coding subspace. Here, multinomial logistic regression identifies coding dimensions that predict one specific condition (colored distributions), versus the other conditions (gray distributions). Because this approach assumes that each neural state is associated with exactly one condition, the last direction vector is fully determined by the rest of the set (i.e. the green axis is the not-blue and not-red axis). In general, whenever there are $k$ exclusive conditions, the coding subspace will have at most $k$-1 dimensions.

**G)** To understand why decoding accuracy improves as we add more neurons, it is helpful to realize that space expands in higher dimensions. Consider the distances between 100 random, uniformly distributed points in 1 or 2 dimensions.

**H)** The expansion means that the distance between distributions will tend to increase as we add more neurons to our decoding model. Compare the difference in coding axis projections as we go from decoding from 1 neuron, to 2 neurons, to 3 neurons.

**I)** Although each new neuron adds noise, neural states within distributions (red trace) will always be closer together than neural states between distributions (gray traces), unless those distributions overlap perfectly. As the dimensionality of the model increases, the distance between distributions grows more rapidly than the distance within distributions. Pairwise correlations between neurons limit information when they cause the dimensionality of the manifold to grow more slowly than the number of recorded neurons. Distances were calculated over 100 simulated trials with 1 to 50 neurons with independent, unit-variance Gaussian noise, at 3 different effect sizes (0.5, 1, 2).

**Text Box 2: SINGLE TRIAL RESOLUTION**

One draw of the population approach is the ability to decode from neural activity finer time scales than is possible with single neurons: the ability to advantage of the "lateral power" we can achieve by combining across neurons, as opposed to the "vertical power" we would achieve through combining across trials (Stokes and Spaak 2016). Population-level analyses have better temporal resolution than single-neuron analyses because adding neurons to an analysis can add new dimensions (Stokes and Spaak 2016; Leavitt et al. 2017; Umakantha et al. 2020). With more dimensions, we can increase the precision with which we can decode continuous information (because we can triangulate decoding across differently tuned neurons) or the reliability with which we can classify trials into their respective conditions (because adding dimensions improves the signal to noise ratio; **Figure 2G-I**).

Neurons do not have to be recorded simultaneously to take advantage of lateral power. Several prominent population neurophysiology studies were actually based on non-simultaneously recorded neurons, which were later combined into *pseudopopulations* (Churchland et al. 2012; Machens et al. 2010; Mante et al. 2013; Meyers et al. 2008). A psuedopopulation is an estimate of what the population response might have been if neurons were recorded simultaneously. It is constructed through bootstrapping: sampling firing rates randomly with replacement from non-simultaneously recorded neurons into new collection of pseudotrials. Although the pseudopopulation approach disrupts the true correlation structure between neurons and is ill-suited to studying events that occur at random times within trials or on a random subset of trials, it can deliver single-trial insight into neural computations. It thus remains a powerful way to do population-level analyses of older datasets or in regions that are not yet amenable to high-yield recording technologies.

Some questions we can only answer with truly simultaneous, high yield recording. The best example within cognitive neuroscience comes from perceptual decision-making. Perceptual decisions about time-varying signals are typically modeled as the slow integration of evidence in favor of one decision or another (Gold and Shadlen 2007). Neural activity mirrors this hypothesized integration process: the firing rates of lateral intraparietal neurons ramp up slowly before the decision (Shadlen and Newsome 2001; Gold and Shadlen 2007). However, because single neurons are averaged over trials, ramping could also be the result of single neurons or the population shifting from some undecided, uncommitted state to a decided state at some random time point on each trial (Latimer et al., 2015; Churchland et al. 2011; Wong and Wang 2006). Combining activity across trials would make activity appear to ramp, even if the underlying generative process was actually discrete steps at random times.

No clear consensus has yet emerged on the stepping-versus-ramping debate in perceptual decision-making (Zoltowski et al. 2019; Shadlen et al. 2016; Chandrasekaran et al. 2018), and the truth is probably more complicated than either simplistic hypothesis (Zoltowski et al. 2019; Daniels, Flack, and Krakauer 2017). However, the new techniques developed over the course of this debate could help answer other important questions in cognitive neuroscience, like if items held in working memory are maintained tonically or juggled dynamically (Lundqvist et al. 2016; E. K. Miller, Lundqvist, and Bastos 2018). These methods could also help us uncover the causes and consequences of the dynamic alternation processes we see when multiple items compete for attention (Engel et al. 2016; Fiebelkorn and Kastner 2019; Caruso et al. 2018) or decision-making (Rich and Wallis 2016), or to understand neural computations that may not be so tightly aligned with task events (Jones et al. 2007; Sadacca et al. 2016; Morcos and Harvey 2016). We will return to some of these ideas when we discuss **Dynamics**.

## SUBSPACES

To make sense of heterogeneity across individual neurons, we have traditionally grouped them into classes. At the population level, we analogously identify *subspaces*. We can think of a subspace as lower dimensional projection of the neural state space: a restricted number of dimensions within the larger state space (Cunningham and Yu 2014). Each neuron's response could be considered a subspace of the population, for example, or a subspace could be composed of some weighted combination of neuronal firing rates across the entire population. Although the term "subspace" is sometimes used loosely in neuroscience, a set of dimensions only make a proper, mathematical subspace if they are orthogonal to each other. Caution is important here because projecting neural activity into non-orthogonal axes can warp the relationships between neural states in unexpected ways.

Often, we are interested in subspaces that encode a piece of information or perform some function. Imagine a neural population that is mixed selective for stimulus color and other information, like location (Rigotti et al. 2013; Raposo et al. 2014). The color subspace of this population would be the portion of the neural space spanned by the color coding dimensions (e.g. red-, green, and blue-coding axes) and the location-subspace would be the portion spanned by location-coding dimensions. These subspaces can be thought of as higher-dimensional generalizations of the coding dimension: neural activity is only "stiff" with respect to the encoded variable within the subspace. Recent work here has looked at the neural subspaces responsible for processing different task dimensions, particularly in perceptual (Mante et al. 2013; Aoi, Mante, and Pillow 2020) and value-based decision-making (Ebitz et al. 2020). This latter study, for example, compared color- and shape-subspace projections while monkeys followed simple color- and shape-based sensorimotor rules (like "choose blue things" or "choose triangles"). It found that information coding in different subspaces was not fixed, but instead was gated by their relevance to the current goal of the animal. This could explain why rule-based decisions were also more energetically efficient than other types of decisions in this study: perhaps rules make efficient use of limited neural resource because using them allows us to collapse the goal-irrelevant dimensions of the neural population code.

One special kind of subspace is a *nullspace*: a subspace that *is not* associated with some function or piece of information (Kaufman et al. 2014). This term was first used to refer to the portion of the neural state space that bears no relationship to motor effectors and thus is likely more involved in the preparatory or cognitive aspects of motor control (Kaufman et al. 2014). We could also imagine a nullspace in which neural activity is not related to stimulus color, location, or the choice the monkey will make (i.e. the dimensions of neural activity that are "sloppy" with respect to these variables). There is a trivial way to create a nullspace (or indeed any other kind of subspace): to have a particular function performed by some segregated group of neurons within the population. If motor effectors were only meaningfully coupled to the spinal-projecting neurons in motor cortex, for example, then it would be trivially true that we would find a motor effector subspace composed only of these neurons. It would also be trivially true that we would have a motor effector nullspace composed of all the other neurons. However, Kaufman et al. (2014) found that this was not the case. The nullspace in this paper was composed of a weighted combinations of neurons, rather than a segregated population. This result strongly resonates with the idea that mixed selectivity is a ubiquitous property of single neurons (Rigotti et al. 2013; Raposo, Kaufman, and Churchland 2014)—an idea that implies that it is the pattern of activity across a population of neurons, rather than a segregated group, that implements some function or encodes some piece of information.

The term subspace can also be used in a way that draws inspiration from neural manifolds, rather than coding dimensions. Not all neural dimensions are needed during each epoch of a task. During a task epoch that does not require working memory, for example, neural dimensions that are responsible for working memory should not be occupied. Thus, by identifying the portions of the neural manifold that are occupied across epochs—the subspaces that define each epoch—we can gain insight into the relationship between the cognitive operations that are occurring (Elsayed et al. 2016). For example, one recent decision-making study compared the neural subspaces across epochs when subjects viewed two offer cues in sequence (Yoo and Hayden 2020). If the neural populations simply encoded offer value, we would expect these subspaces to be identical. However, the second offer was encoded in a different subspace from the first, so that the evaluation and comparison steps were functionally – and not anatomically - segregated

We have seen that subspaces can be defined based on their relationship to task variables, epochs, and motor effectors (Kaufman et al. 2014), but they can also be defined by their relationship to other brain regions (Semedo et al. 2019). Here, instead of using regression to identify a coding dimension that maps neural activity to some task dimension, researchers identify dimensions in one neural population that explain variability in a second population. This approach has the potential to be incredibly powerful because it can isolate the information shared by two regions in the *communication subspace* (Semedo et al. 2019), the private information within each region's nullspace, and determine how this information changes over time (Perich et al. 2020). However, these methods have the same limitations as all correlational methods: they do assume some directionality but cannot uniquely determine how information flows between regions or whether it is inherited from some third region that influences both.

## DYNAMICS

Much of the work we have talked about so far assumes that neural activity can be summarized by a single neural state, a single point in time. However, neural activity clearly evolves over time and across neurons in important ways. A single-neuron neurophysiologist might examine these patterns through scrolling through the PSTHs of the individual neurons or averaging them all together. At the population level, we consider an entire collection of PSTHs at the same time through examining neural trajectories: the paths activity traces through the neural state space (**Figure 1B**). Why do neural trajectories evolve as they do? One influential idea is that trajectories are shaped by hidden network-level forces called *dynamics* (**Figure 3A**). If we imagine neural activity as a ball rolling across a landscape, the trajectory is the path traced by the ball, but the dynamics are the forces that shape its path: the landscape itself. We generally cannot measure dynamics because they are a complex function of features we cannot resolve experimentally: the networks' wiring, inputs and activation functions (though see (Genkin, Hughes, and Engel 2020)). However, we can make inferences about dynamics by looking at the frequency of different neural states, examining the shape of neural trajectories, comparing data with computational models, or looking at how the network responds to perturbations. There are many excellent reviews on population dynamics (Brody, Romo, and Kepecs 2003; Vyas et al. 2020; Sussillo 2014; Chaudhuri and Fiete 2016; Shenoy and Kao 2021; Yuste 2015; Ju and Bassett 2020), so we will focus on their links to cognitive processes here.

Perhaps the most fundamental concept in population dynamics is the *attractor*. An attractor is a valley in the landscape, a neural state (or collection thereof) towards which nearby activity will evolve. Many neural network architectures naturally produce multiple stable, *fixed point* attractors and theoretical work has long implicated these in cognition (Moreno-Bote,

Rinzel, and Rubin 2007; Hopfield 1982; Miller 2016; Wong and Wang 2006). For example, a Hopfield network (an early RNN) shows how a system of attractors can solve important problems in associative memory (Hopfield 1982). Because they pull nearby patterns of activity towards a stable state, attractors are ideally suited to solve cognitive problems like categorizing an input or holding an item stably in memory (Miller 2016).

When a network includes multiple attractors, we can see evidence of "multi-stability" in its activity (Miller 2016): the network transitions between essentially discrete neural states (**Figure 3B**). Theoretical work suggests that this is a natural consequence of clustered connections between neurons (Litwin-Kumar and Doiron 2012). In neurophysiology data, multi-stable patterns often appear when multiple objects are in competition, be it for visual (Engel et al. 2016; Fiebelkorn and Kastner 2019) or auditory attention (Caruso et al. 2018), or in value-based decision-making (Rich and Wallis 2016; Hayden and Moreno-Bote 2018). For example, one visual attention study reported multi-stable population states within cortical columns in visual cortex (Engel et al. 2016): neurons selective for one stimulus were either on (highly active), or off (largely inactive), a striking phenomenon observable only at the population level. Attention regulated the duration of on-states, which, in turn, predicted the detection of subtle stimulus changes and may even explain some single-neuron correlates of selective attention. In this study, the transitions between states were stochastic, but attention can also be reoriented rhythmically (Dugué, Roberts, and Carrasco 2016; Fiebelkorn and Kastner 2019) and future work is needed to understand whether transitions between multi-stable attractors are driven rhythmically or stochastically.

Sequences of multi-stable attractors could implement computations (Miller 2016; Vyas et al. 2020; Sussillo 2014; Neves and Timme 2012). For example, the classic attractor network model of decision-making evolves from an undecided state to one out of several possible absorbing, "decided" states over time, with evidence (Wong and Wang 2006; Wang 2008). This style of model predicts that perturbing neural activity should have a larger effect early on in the process, before the system has settled into a decided state, than it does later the vicinity of the attractor, where stabilizing forces are greater. Indeed, this is exactly what is seen experimentally with optogenetic perturbations in rodent decision-making regions (Kopec et al. 2015).

Of course, these models are highly simplified. They can be implemented in a network containing only 2-3 neuronal pools and not all decision-making is best made by ballistic, winner-take-all dynamics. In a more naturalistic, consummatory decision-making paradigm, animals transitioned between sequences of 4-5 discrete neural states as they evaluated and decided whether to accept or reject a proffered reward (Sadacca et al. 2016). Other more complex neural network models cast perceptual decision-making as the product of a line attractor (Sussillo 2014; Mante et al. 2013), meaning a set of multi-stable attractors that neural activity is driven along by each new piece of evidence. Different regions may also implement different dynamics in the same task. For example, in rats performing an evidence accumulation task, researchers found evidence of graded, line-attractor-like representations in posterior cortex, but ballistic, winner-take all dynamics in frontal cortex (Hanks et al. 2015)—despite nearly identical averaged PSTHs in the two regions.

We have considered only stable, fixed point attractors, but attractors are not necessarily stable in all dimensions, especially in high-dimensional systems. Activity may be pulled into an attractor from one direction, only to be pushed away, towards another attractor, along another direction. Collections of this type of partially stable attractor are known as heteroclinic channels and the trajectories they produce are known as heteroclinic cycles or orbits (Neves and Timme 2012; Rabinovich et al. 2001; Rabinovich, Huerta, and Laurent 2008), though they can have any shape. In neuroscience, the term *dynamic attractor* is perhaps more common (Laje and Buonomano 2013). Dynamic attractors can produce complex, continuously evolving trajectories through the neural state space and are, in theory, capable of implementing any arbitrary computation (Neves and Timme 2012).

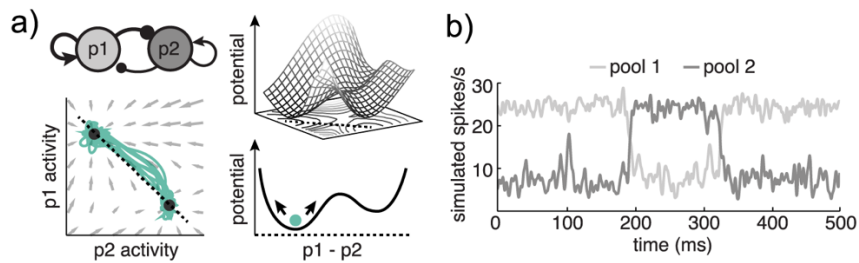

**Figure 3: Dynamics**

**A)** A 2-pool neural network (similar to Figure 2B) and 3 different views of the network's dynamics. Bottom left) A *phase portrait* shows the direction and magnitude of the local forces in the network (gray arrows). Simulated trajectories are overlaid (pale green traces). Filled circles are the fixed points at the center of the attractors' basins. Top right) We can also visualize the dynamics as a potential energy landscape, which highlights the unstable peaks and stable valleys that shape how activity evolves over time. Bottom right) A cartoon illustrating one slice through the potential energy landscape, following the typical path of trajectories in the state space (i.e. the dotted line in the phase portrait).

**B)** One random simulation of the network in (A), illustrating the activity in each pool of neurons as a function of time. Noise is sufficient to cause the network to hop from one stable state (p1 > p2) to a second (p2 > p1) and back again. Over many simulations, the duration of time spent in each state will be proportional to the relative depth of the states in (A).

## OPEN QUESTIONS

Population neurophysiology has its own object of study, characteristic set of methods, and suite of key concepts that give us new ways to reason about how neurons behave collectively, rather than as individuals. We have introduced 5 of these concepts here: the *neural states* that provide a snapshot of a pattern of activity across the population, the *manifold* that encompasses the neural states that are possible (Manifold) or at least observed (manifold), the *coding dimensions* and *subspaces* that link neural states to behavior and cognition, and the *dynamics* that map activity from neural state to neural state, guiding how trajectories evolve through time and across the state space. These concepts have links to ideas and analyses from the single-neuron approach—links which we have worked to highlight here—but despite our best efforts, these mappings are not perfect. There is something new about population-level thinking that simply cannot be understood as a composition of single-neuron concepts, just as the population itself cannot be understood as a composition of single neurons. To us, these are signs that population neurophysiology is coalescing into a new field. However, important conceptual and methodological questions remain.

Conceptually, we should acknowledge that the neural population doctrine has a weakness that is not shared with the single-neuron doctrine. The limits of a neuron are obvious—it has cell walls—but what are the limits of a population? Are its boundaries the set of recorded neurons? The tissue surrounding the electrodes? The edges of the Broadman area? The skull? Advances in neural recording technologies may move us towards this broadest notion of a neural population, but, for now, the term is ambiguous. It is not always immediately clear if a paper shares our notion of population, or when the term population is distinct from related terms, like "neuronal ensembles". It may not be necessary to have a formal definition, provided practitioners of the population approach are able to understand each other, but the lack of a definition does present opportunities for confusion.

Methodologically, it is not clear that we have fully come to terms with some of the implications of the population doctrine. If our goal is to understand how neural populations behave collectively, is spike sorting still necessary or does it become irrelevant? Spike sorting is an imperfect process, and, despite concerted efforts to automate it, remains incredibly time consuming. This bottleneck that will only grow as neuronal yields continue to accelerate. Though state-of-the art algorithms perform well, they require experimentalists to sacrifice the spatial extent of recordings for dense, overlapping coverage of a smaller regions (Rossant et al. 2016). However, isolating individual neurons is probably not critical for every population-level analysis (Trautmann et al. 2019). Developing a better understanding of when isolated cells matter and when they do not could help us move to a model where neurons are isolated only when scientifically necessary.

We are only just starting to explore how this new generation of population-level analyses may connect with other population-level phenomena, like neuronal correlations and field potentials. It is clear that correlations between neurons have substantial, structuring effects on population-level representations (Umakantha et al. 2020; Elsayed and Cunningham 2017). In our view, continued work on these correlations will be an essential bridge between single-neuron and population-level accounts in the future. Local field potentials are emergent, population-level phenomena in their own right. Although we have not yet been explicit on this point, oscillations across populations of neurons can be used to index population computations in ways that can either exactly parallel spiking results or deliver important new insights (Hunt et al. 2015; Chaudhuri et al. 2019; Gallego-Carracedo et al. 2021; Hall, de Carvalho, and Jackson 2014; Lundqvist et al. 2016; Smith et al. 2019; Widge et al. 2019). If population neurophysiology continues to evolve towards a broader notion of the limits of a population, perhaps we will see more researchers embrace the insights that are currently only possible with broad scale electrophysiological measures like local field potentials, electrocorticography, electroencephalography, and/or magnetoencephalography.

## CONCLUSIONS

Some scholars have argued that it is time to replace the single-neuron doctrine with the neural population doctrine (Saxena and Cunningham 2019). We are sympathetic to this view. So was Canadian neuroscientist Donald Hebb, who argued that the assembly—a collection of cells wired up together—should be considered the fundamental unit of computation because it produces reverberating behavior that cannot be understood from isolating each neuron (Hebb 1949). Theoretical neuroscientists have long worked at the level of the population, extrapolating insights from single neurons into population-level computational models (Laje and Buonomano 2013; Hopfield 1982; Wong and Wang 2006; Bernardi et al. 2020; Moreno-Bote et al. 2007; Chaudhuri and Fiete 2016; Brody et al. 2003; Miller 2016; Sussillo 2014; Yuste 2015; Litwin-Kumar and Doiron, 2012). Even some traditional physiological techniques, like the population average, can reveal population-level phenomena, even when not explicitly recognized as a population-level analyses. Clearly, then, this view itself is not new. What is new is the ability to readily record from large numbers of neurons, the development of new tools to analyze population activity, and a suite of concepts that constitute a systematic new framework for reasoning about neural activity at the population level.

Some have argued that population-level phenomena must be *emergent*, that is, exist only at the population level, to be interesting (Saxena and Cunningham 2019; Pillow and Aoi 2017; Elsayed and Cunningham 2017). Focusing on emergent phenomena may be the most straightforward way to overcome the occasional misconception that population neurophysiology is a trivial repackaging of what we already knew, only now with more neurons. However, it is not necessary for a phenomenon to exist only at one level to be relevant to that level. Even if a population-level result could be entirely predicted from the properties of single neurons, did anyone make the prediction? Does the population-level description give us a new hypothesis to test, a more compact description, or a surprising new view? We must remain cautious in our armchair intuitions about how single-neuron results will map to the population level. We almost certainly could have intuited the existence of coding dimensions from single-neuron tuning, but our intuition about how tuned neurons contribute to coding dimensions would have been wrong (Leavitt et al. 2017), we probably would not correctly predict coding dimension geometry (Okazawa et al. 2021), and we may not have even considered that coding dimensions may be dynamic (Stokes et al. 2013). Translating ideas in new frameworks is rarely trivial, even if it may take a few years to know whether a population-level explanation is completely novel or just a critical bridge between increasingly disparate levels of analysis.

**Acknowlegements.** This work was supported by the Natural Sciences and Engineering Research Council (RBE: Discovery Grant RGPIN-2020-05577), the Fonds de la Recherche en Santé du Québec (RBE: Junior 1 Salary Award #284309), the Brain and Behavior Research Foundation (RBE: Young Investigator Grant #27298), and the National Institutes of Health (BYH: NIDA #038615 and NINDS #118366 ). We would like to thank Katarzyna Jurewicz, Alex Herman, Vince McGinty, Justin Fine, Pouya Bashivan, Suresh Krishna, and three anonymous reviewers for helpful comments on previous versions of this manuscript.